\documentclass[aps,prd,12pt,preprint,tightenlines,
amsfonts,amssymb,amsmath,byrevtex,showpacs]{revtex4}
\begin{document}
\newcommand{\dR}{\mathbb R}
\newcommand{\dC}{\mathbb C}
\newcommand{\dS}{\mathbb S}
\newcommand{\dZ}{\mathbb Z}
\newcommand{\id}{\mathbb I}
\newcommand{\ep}{\epsilon}
\newcommand{\be}{\begin{equation}}
\newcommand{\ee}{\end{equation}}

\title{Quantum particle on hyperboloid}

\author{W\l odzimierz Piechocki}
\affiliation{So\l tan Institute for Nuclear Studies, Ho\.{z}a 69,
00-681 Warszawa, Poland; e-mail: piech@fuw.edu.pl}

\begin{abstract}
We present quantization of particle dynamics on one-sheet hyperboloid 
embedded in three dimensional Minkowski space. Taking account of all 
global symmetries of the system enables unique quantization. Making use of 
topology of canonical variables not only simplifies calculations but also 
gives proper framework for analysis.
\end{abstract}
\maketitle

\section{Introduction}

It is known that canonical quantization of a system with non-trivial topology 
of its phase space (i.e. different from $\dR^{2n}$) is a nonunique procedure 
(see, e.g. \cite{CI,BS}). It has been found  that ambiguities of 
quantization may be reduced by taking into account global properties of 
a system \cite{PLB,WP}.

In \cite{WP} we have quantized the dynamics of a relativistic test particle 
on a one-sheet hyperboloid embedded in three dimensional Minkowski space, 
i.e. on the two dimensional de Sitter space with the topology $\dR\times\dS$. 
The resulting quantum theory depends on a real parameter $\theta$. To simplify 
the discussion we fixed the parameter to its values $\theta =0,$ 
which corresponds to the choice $SO_0(1,2)$ as the symmetry group of the 
system. In the present work we discuss the $\theta-$dependence of the results. 
We show that making use of time-reversal invariance of the system fixes 
the parameter to its two values: $\theta =0$ and $\theta = \pi.$ 

In our quantization procedure we make use of the idea introduced in 
\cite{KRP} of taking $U(\beta):=\exp(i\beta),~0\leq\beta <2\pi,~$ to represent 
the variable with topology $\dS$, instead of the common choice 
$\tilde{U}(\beta):=\beta$. This idea has proved to be fruitful in  the
quantization of particle dynamics on a circle \cite{KRP} and on a sphere 
\cite{KR}. 

\section{Observables}

The system of a relativistic test particle on a one-sheet hyperboloid 
\be
(y^2)^2 + (y^1)^2 - (y^0)^2 =r^2_0,
\ee
where $y^a ~(a=0,1,2)$ denote coordinates of 3-dimensional Minkowski space 
and $r_0$ is the parameter specifying the hyperboloid,
has been found to be integrable \cite{GW}. 

The space of timelike geodesics of a particle may be determined by the 
solutions to the algebraic equations \cite {GW} 
\be
J_a y^a =0,~~~~J_2 y^1 - J_1 y^2 =r_0 p,
\ee
where $p$ denotes one of the two canonical momenta of a particle. 
The dynamical integrals $J_a ~(a=0,1,2)$, owing to the constrained dynamics 
of the system, satisfy the equation  \cite{GW}
\be
J_2^2 +J_1^2 -J_0^2 = \kappa^2,~~~~\kappa:= m_0 r_0,
\ee
where $m_0$ denotes particle's mass. $J_a~ (a=0,1,2)$ are 
generators of the proper orthochronous Lorentz group $SO_0(1,2)$; $J_0$ 
corresponds to the invariance of (1) with respect to rotations, whereas $J_1$ 
and $J_2$ describe two boosts. The generators of $SO_0(1,2)$ group 
are basic observables of the system \cite{WP}. 

The phase space $\Gamma$ of the system is defined to be the space of all 
timelike geodesics of a particle available for dynamics. Each point 
$(J_0,J_1,J_2)$ of the hyperboloid (3) specifies a geodesic with coordinates 
satisfying (1) and (2). Thus, the hyperboloid (3) plays 
the role of $\Gamma$. The system has two degrees of freedom and $\Gamma$ 
(in what follows being identified with (3)) is globally homeomorphic to the 
canonical phase space $X$ defined as
\be
X:=\{(J,\beta)~|~J\in \dR,~~0\leq \beta <2\pi \}, 
\ee
where the homeomorphism may be defined \cite{WP} by
\be
J_0 =J,~~~J_1 =J\cos\beta -\kappa\sin\beta,
~~~J_2 = J\sin\beta +\kappa\cos\beta.
\ee
Introducing the Poisson bracket on $X$ by 
\be
\{ \cdot, \cdot\}:= \frac{\partial\cdot}{\partial J}\frac{\partial\cdot}
{\partial \beta}- \frac{\partial\cdot}{\partial \beta}\frac{\partial\cdot}
{\partial J},  
\ee
we obtain 
\be
\{ J_0,J_1\}= -J_2,~~~\{ J_0,J_2\}= J_1,~~~\{ J_1, J_2\}=J_0,
\ee
which means that the observables $J_a~(a=0,1,2)$ satisfy the algebra 
$so(1,2)$. The algebra (7) describes local symmetry of the physical phase 
space (3).  Since the topology of the hyperboloid (3) is $\dR \times\dS,$ 
the global symmetry of $\Gamma$ may be taken to be any group with $so(1,2)$ 
as its Lie algebra, i.e. $SO_0(1,2)$ or any covering of it. 

Untill now we have considered only  \textit{continuous} transformations, 
but our system may 
be also invariant under \textit{discrete} transformations. Since the 
system of a particle on hyperboloid is a non-dissipative one, 
it must be invariant with respect to time-reversal transformations $T$. 
We postpone further discussion of $T-$invariance to the section dealing 
with quantization. 

\section{Redefinitions}

It is advantagous, for further considerations, to redefine the basic 
observables (5) and the canonical variables (4) as follows
\be
J_0 :=J,~~~J_+ :=J_1 +iJ_2 =(J+i\kappa)U,~~~J_- :=J_1-iJ_2 =(J-i\kappa)U,
\ee
where $U:=\exp(i\beta),~~0\leq \beta <2\pi$.

\noindent
The canonical phase space $X$ is now represented by 
\be
X=\{ (J,U)~|~J\in\dR,~U\in\dS\}.
\ee
It has been shown in \cite{KRP} that making use of $U$, instead of 
$\beta$, to represent the variable with $\dS$ topology is mathematically 
better justified.

We also redefine the algebra multiplication replaceing (6) by 
\be
\ll\cdot,\cdot \gg:= (\frac{\partial\cdot}{\partial J}\frac{\partial\cdot}
{\partial U}- \frac{\partial\cdot}{\partial U}\frac{\partial\cdot}
{\partial J})U =\{\cdot,\cdot\}U.
\ee
One can check that owing to the Poisson bracket properties the bracket (10) 
satisfies the three axioms: linearity, antisymmetry and the Jacobi identity. 
Thus (10) defines the Lie multiplication.

One can easily verify that the algebra (7) in new disguise reads
\be
\ll J_0,J_+\gg =J_+,~~~~\ll J_0,J_-\gg =-J_-,~~~~\ll J_-,J_+\gg =2J_0.
\ee
In particular, the new canonical variables satisfy the algebra
\be
\ll J,U\gg =U
\ee
to be compared with the algebra
\be
\{J,\beta\}=1
\ee
 based on (4) and (6).
 
 \section{Quantization}
 
 By quantization of particle dynamics we mean finding an (essentially) 
 self-adjoint representation of the algebra of observables integrable 
 to an irreducible unitary representation of the symmetry group of the 
 system. The representation space plays the role of the quantum states space.
 
 To take into account the time-reversal invariance of the system we require 
 the representation algebra to be invariant under time-reversal operator 
 representing time-reversal transformations.
 
 \subsection{Representation of canonical variables algebra}
 
 To quantize the algebra (11), we first quantize the canonical variables 
 algebra (12). Following the idea presented in \cite{KRP} we make use 
 of the mapping
 \be
 J\rightarrow \hat{J}\psi(\beta):= -i\frac{d}{d\beta}\psi(\beta),~~~~~
 U=e^{i\beta}\rightarrow \hat{U}\psi(\beta):= e^{i\hat{\beta}}\psi(\beta):= 
 e^{i\beta}\psi(\beta),
 \ee
 where $\psi\in L^2(\dS)$.
 
 \noindent
 It is easy to check (see, App. A of \cite{WP}) that the operator $\hat{J}$ 
 is essentially self-adjoint on the dense subspace 
 $\Omega_\theta,~0\leq\theta < 2\pi,$ of $L^2(\dS)$ defined to be
 \be
 \Omega_\theta:=\{ \psi\in L^2(\dS)~|~\psi\in C^\infty [0,2\pi],~
 \psi^{(n)}(2\pi)=e^{i\theta}\psi^{(n)}(0),~n=0,1,2,... \}.
 \ee
 It follows easily that the spectrum of $\hat{J}$ reads 
 \be
 sp \hat{J}(\theta)=\{ j:=m+\frac{\theta}{2\pi}~|~m\in\dZ,~0\leq\theta <2\pi\}, 
 \ee
 and $\Omega_\theta$ is spanned by 
 \be
 f_{m,\theta}(\beta)=\frac{1}{2\pi}e^{ i\beta(m+\frac{\theta}{2\pi})}.
 \ee
 We see at once that 
 \be
 [\hat{J},\hat{U}]\psi := \hat{J}\hat{U}\psi -\hat{U}\hat{J}\psi = 
 \widehat{\ll J,U\gg}\psi =\hat{U}\psi,~~~~\psi\in\Omega_\theta ,
 \ee
 since the unitary operator $\hat{U}$ is well defined on the entire 
 $L^2(\dS)$. Therefore, the mapping (14) leads to the essential 
 self-adjoint representation of (12) on $\Omega_\theta$. 
 Since the parameter $\theta$ is a real number, the algebra (12) has 
 an infinite number of unitarily nonequivalent representations. 
 
 \subsection{Representation of observables}
 
 Applying (14) and the symmetrisation prescription to the products $JU$ 
 in (8) we obtain the following mapping
 \be
 J_0\rightarrow \hat{J}_0 :=\hat{J},~~~J_-\rightarrow\hat{J}_- :=\hat{U}^{-1}
 (\hat{J}- 1/2-i\kappa),~~~J_+\rightarrow\hat{J}_+ 
 :=(\hat{J}-1/2+i\kappa)\hat{U},
 \ee
 where $\hat{U}^{-1} :=\exp(-i\hat{\beta}).$
 
 One can easily verify that (19) is a homomorphism
 \be
[\hat{J}_0,\hat{J}_+] = \widehat{\ll J_0,J_+ \gg}=\hat{J}_+,~~~
[\hat{J}_0,\hat{J}_-] = \widehat{\ll J_0,J_- \gg}=-\hat{J}_-,~~~
[\hat{J}_-,\hat{J}_+] = \widehat{\ll J_-,J_+ \gg}=2\hat{J}_0 .
 \ee
The operators (19) and the equations (20) are well defined on 
$\Omega_\theta$. 

Making use of the method applied in App. A of \cite{WP} it is 
straightforward to prove that (19) defines the essentially self-adjoint 
representation of the algebra (11), if the common domain of the algebra 
(20) coincides with $\Omega_\theta$. The proof rests heavily on the fact 
that  on $\Omega_\theta$ the problem reduces to the problem of 
self-adjointness of the representation algebra (18).

\subsection{Time-reversal invariance}

We impose the time-reversal invariance (see, e.g. \cite{WKT}) upon the system 
by the requirement of time-reversal invariance of the algebra (20). 
In what follows we show that (20) is time-reversal invariant, if the algebra 
of canonical variables (18) has this property.

The observable $J$ being associated with the invariance of (1) with respect 
to rotations may be given the interpretation of angular momentum of 
a test particle. Thus, the operator $\hat{J}$ transforms as 
\be
\hat{T}\hat{J}\hat{T}^{-1} =-\hat{J},
\ee 
where $\hat{T}$ is an anti-unitary operator of time inversion.

\noindent
It is easy to check that the algebra (18) is time-reversal invariant 
\cite{KRP}, if the operator $\hat{U}$ transforms  as
\be
\hat{T}\hat{U}\hat{T}^{-1}=\hat{U}^{-1}.
\ee
Owing to (21), (22) and (19) it is easy to check that 
\be
\hat{T}\hat{J}_+ \hat{T}^{-1}=- \hat{U}^{-1}\hat{J}_+\hat{U}^{-1},~~~~~
\hat{T}\hat{J}_- \hat{T}^{-1}=- \hat{U}\hat{J}_-\hat{U}.
\ee
Making use of (23) one can easily show that the algebra (20) is time-reversal 
invariant.
 
Now, let us examine the consequences of the imposition of time-reversal 
invariance in the context of the representation space. It was shown in 
\cite{KRP} that the eigenvalue equation
\be
\hat{J}|j> = j|j>
\ee
combined with (18) leads to 
\be
\hat{U}|j> = |j+1>,~~~~~\hat{U}^{-1}|j> = |j-1>,
\ee
which means that the set of all eigenfunctions $\{|j>\}$ can be generated 
from the 
`vacuum' state $|j_0>$, where $0\leq j_0 \leq 1$. It is a simple matter 
to show that the consistency of Eqs. (21), (22) and (25) yields \cite{KRP} 
\be
\hat{T}|j> = |-j>.
\ee
Therefore, the spectrum of $\hat{J}$ must be symmetric with respect to 
$j$, which means that $j_0 =0$ or $j_0 =1/2$. Hence, $j$ can be only 
integer or half-integer. Comparing this result with (16) and (17), we 
conclude that the range of the parameter $\theta$ must be restricted 
to the two values: $\theta =0$ and $\theta =\pi$.

This way we have proved that the algebra (11) has two classes of 
representations. They are labeled by $\theta =0$ and $\theta =\pi$.

\subsection{Identification of representations}

For clarity, we recall the existence of the following algebra isomorphisms 
\cite{BR}: $~so(1,2)\sim sl(2,\dR)\sim su(1,1),$ and group isomorphisms 
\cite{NJV,PJS}:
$~SL(2,\dR)\sim SU(1,1)$ and $~SO_0(1,2)\sim SL(2,\dR)/\dZ_2 \sim
\widetilde{SL(2,\dR)/\dZ}$.

To compare our representations with known representations of the Lie groups 
having mentioned above isomorphic Lie algebras, we consider the Casimir 
operator of the algebra (11). Its classical and quantum versions, 
respectively, read 
\cite{WP} 
\be
C:= J_2^2 +J_1^2 -J_0^2=J_+ J_- -J^2 =\kappa^2,~~~~\hat{C}:=
\frac{1}{2}(\hat{J}_+\hat{J}_- + 
\hat{J}_- \hat{J}_+)- \hat{J}^2= (\kappa^2 + 1/4)\id.
\ee
One can check at once that $[\hat{C},\hat{J}]=0$, which means that the 
representation space may be labelled by the three parameters: 
$\theta$, $\kappa$ and $j$.

Since $1/4 < \kappa < \infty$  the class with $\theta =0$ 
and $j$ integer, and the class with $\theta =\pi$ and $j$ half-integer 
belong to the principal series single valued and double valued, 
respectively, unitary representation of $SU(1,1)$ group \cite{VB,NJV}. 
The single valued case coincides with the representation of $SO_0(1,2)$ 
group \cite{VB}. 

Therefore, the principal series representation of $SL(2,\dR)$ 
group unifies the principal series representation of $SO_0(1,2)$ group 
and the representation of the time-reversal transformations. 

The general case $0\leq\theta < 2\pi$  corresponds \cite{PJS} to the principal 
series representation of the universal covering group $\widetilde{SL(2,\dR)}$.

\section{Conclusions}

Quantization based on \textit{local} properties  of a classical system only,
leads to nonuniqueness. We have seen that quantization of the local symmetry 
without the requirement of its integrability to the global symmetry yields 
infinitely many quantum systems corresponding to a single classical system. 
Such approach is also not suitable for quantization of systems with singular 
spacetimes \cite{PLB,WP}. \textit{Global} symmetries (continuous and discrete)  
turn out to be of primary importance. 
Making use of them we have managed to avoid problems in quantum theory 
connected with removable type singularities of spacetime  \cite{WP} and have  
succeeded to get  unique results.  Obviously, we have not 
discussed here the commonly known ambiguity problem (see, e.g. \cite{RL}) 
connected with the imposition of quantum rules on constrained 
dynamics (quantize first then impose constraints or vice versa problem). 
But this problem is beyound the scope of the present paper.

We have found that the problem of self-adjointness of observables and 
time-reversal invariance of the system may be already implemented at the 
level of \textit{canonical variables} algebra. This is possible owing 
to the choice of canonical variables in the form which fits the 
\textit{topology} of the variables. The common choice of variables on $\dS$ 
in the form used for an interval $[a,b]\subset \dR$, makes impossible the 
solution of the above problems at the canonical level. It is so because one 
cannot find a self-adjoint representation of the algebra 
(see, e.g. App. B of \cite{WWW}). It may be solved, but at the level of the 
observables algebra (see, e.g. App. A of \cite{WP}). The choice of the 
canonical variables in the form compatible with their topologies seems 
to have basic significance. 
Finally, we make comment concerning our choice of the Lie multiplication 
(10). Comparing (18) with (12), and (20) with (11) we cannot see the factor 
$(-i)$, which usually occurs in homomorphisms from classical to quantum 
algebras. This is a nice feature of (10) and it is again connected with 
the topology consistent form of canonical variables.

\begin{acknowledgments}
The author would like to thank Professor J. Rembieli\'{n}ski for very fruitful 
suggestions concerning the solution of the ambiguity problem.
\end{acknowledgments}

\end{document}